\documentclass[reprint,
 superscriptaddress,
 amsmath,amssymb,
 aps,
 prl
]{revtex4-2}

	\usepackage[T1]{fontenc}
	\usepackage[utf8]{inputenc}
    \usepackage[pdftex,bookmarks=false]{hyperref}
    \usepackage{microtype}
    \usepackage{graphicx}
    \usepackage{dcolumn}
    \usepackage{bm}
    \usepackage{multirow}
    \usepackage{booktabs}
    \usepackage{xcolor}
    \usepackage{mathtools}
    \usepackage{mathptmx}
    \usepackage{upgreek}
    \usepackage{float}

    \hypersetup{
    	colorlinks = true, 
		citecolor = blue,
		bookmarksnumbered = true,
		urlcolor = green
	}
	
    \DeclarePairedDelimiter\bra{\langle}{\rvert}
    \DeclarePairedDelimiter\ket{\lvert}{\rangle}
    \DeclarePairedDelimiterX\braket[2]{\langle}{\rangle}{#1 \delimsize\vert #2}

    \renewcommand{\Re}{\operatorname{Re}}
    \renewcommand{\Im}{\operatorname{Im}}
    \newcommand{\beginsupplement}{
        \setcounter{table}{0}
        \renewcommand{\thetable}{S\arabic{table}}
        \setcounter{figure}{0}
        \renewcommand{\thefigure}{S\arabic{figure}}
        \setcounter{equation}{0}
        \renewcommand{\theequation}{S\arabic{equation}}
     }

\begin{document}
\title{Nodal line resonance generating the giant anomalous Hall effect of Co$_3$Sn$_2$S$_2$}
\date{\today}

\author{F.~Schilberth}
\affiliation{Experimentalphysik V, Center for Electronic Correlations and Magnetism, Institute for Physics, Augsburg University, D-86135 Augsburg, Germany} 
\affiliation{Department of Physics, Institute of Physics, Budapest University of Technology and Economics, M\H{u}egyetem rkp. 3., H-1111 Budapest, Hungary}

\author{M.-C. Jiang}
\affiliation{Department of Physics and Center for Theoretical Physics, National Taiwan University, Taipei 10617, Taiwan}
\affiliation{RIKEN Center for Emergent Matter Science, 2-1 Hirosawa, Wako 351-0198, Japan}

\author{S. Minami}
\affiliation{Department of Mechanical Engineering and Science, Kyoto University, Nishikyo-ku, Kyoto 615-8540, Japan}
\affiliation{Department of Physics, University of Tokyo, Bunkyo-ku, Tokyo 113-0033, Japan}

\author{M.~A.~Kassem}
\affiliation{Department of Materials Science and Engineering, Kyoto University, Kyoto 606-8501, Japan} 

\author{F.~Mayr}
\affiliation{Experimentalphysik V, Center for Electronic Correlations and Magnetism, Institute for Physics, Augsburg University, D-86135 Augsburg, Germany}

\author{J.~Deisenhofer}
\affiliation{Experimentalphysik V, Center for Electronic Correlations and Magnetism, Institute for Physics, Augsburg University, D-86135 Augsburg, Germany}

\author{T. Koretsune}
\affiliation{Department of Physics, Tohoku University, Sendai 980-8578, Japan}

\author{Y.~Tabata}
\affiliation{Department of Materials Science and Engineering, Kyoto University, Kyoto 606-8501, Japan} 

\author{T.~Waki}
\affiliation{Department of Materials Science and Engineering, Kyoto University, Kyoto 606-8501, Japan} 

\author{H.~Nakamura}
\affiliation{Department of Materials Science and Engineering, Kyoto University, Kyoto 606-8501, Japan} 

\author{G.-Y. Guo}
\affiliation{Department of Physics and Center for Theoretical Physics, National Taiwan University, Taipei 10617, Taiwan}
\affiliation{Physics Division, National Center for Theoretical Sciences, Taipei 10617, Taiwan}

\author{R. Arita}
\affiliation{RIKEN Center for Emergent Matter Science, 2-1 Hirosawa, Wako 351-0198, Japan}
\affiliation{Research Center for Advanced Science and Technology, University of Tokyo, 4-6-1 Meguro-ku, Tokyo, 153-8904, Japan}

\author{I.~K\'ezsm\'arki}
\affiliation{Experimentalphysik V, Center for Electronic Correlations and Magnetism, Institute for Physics, Augsburg University, D-86135 Augsburg, Germany}

\author{S.~Bord\'acs}
\affiliation{Department of Physics, Institute of Physics, Budapest University of Technology and Economics, M\H{u}egyetem rkp. 3., H-1111 Budapest, Hungary}
\email{bordacs.sandor@ttk.bme.hu}

\begin{abstract}
Giant anomalous Hall effect (AHE) and magneto-optical activity can emerge in magnets with topologically non-trivial degeneracies. However, identifying the specific band structure features like Weyl points, nodal lines or planes which generate the anomalous response is a challenging issue. Since the low-energy interband transitions can govern the static AHE, we addressed this question in the prototypical magnetic Weyl semimetal Co$_3$Sn$_2$S$_2$ also hosting nodal lines by broadband polarized reflectivity and magneto-optical Kerr effect spectroscopy with a focus on the far-infrared range. In the linear dichroism spectrum we observe a strong resonance at 40\,meV, which also shows up in the optical Hall conductivity spectrum and primarily determines the static AHE, thus, confirms its intrinsic origin. Our material-specific theory reproduces the experimental data remarkably well and shows that strongly tilted nodal line segments around the Fermi energy generate the resonance. While the Weyl points only give vanishing contributions, these segments of the nodal lines gapped by the spin-orbit coupling dominate the low-energy optical response. 
\end{abstract}
\maketitle


Topological Dirac and Weyl semimetals have received much attention, since at low-energies their electrons mimic relativistic particles \cite{Armitage2018}. Moreover, topological semimetals with higher dimensional degenerate manifolds, such as nodal lines and even planes, have also been predicted and observed, which host quasiparticles that are unprecedented in particle physics \cite{Burkov2011a,Bradlyn2016,Bzdusek2016,Wilde2021}. These peculiar band structure features give rise to e.g.~exceptionally high mobility \cite{Liang2015,Shekhar2015}, chiral anomaly \cite{Xiong2015,Huang2015}, Fermi arcs and drumhead surface states \cite{Liu2014,Liu2014a,Xu2015,Belopolski2019}, and unusual quantization of orbital motion in a magnetic field \cite{Yuan2018,Zhao2022}.

Recently, the search for such topological band features in magnetic materials has become a hot topic. In magnets, the topological nodes can be controlled by magnetic fields \cite{Chang2018,Yin2018,Li2019}, they induce exotic domain wall states \cite{Ueda2014, Destraz2020} and generate enhanced anomalous Hall effect (AHE) \cite{Burkov2011,Armitage2018}. The intrinsic AHE being proportional to the Berry-curvature integrated over the Brillouin zone (BZ) has particular importance as it is a direct consequence of the non-trivial band topology \cite{Nagaosa2010}. At the heart of these phenomena, there is the interplay between the magnetic order and the band structure mediated by the spin-orbit coupling (SOC). In the ordered state, the broken spin-rotation symmetry may either reduce the degeneracy of the manifolds, e.g.~transform a nodal line into Weyl points \cite{Liu2018,Fang2016}, or completely gap out the nodes, which may stabilize a topological insulator phase \cite{Yang2018}. Therefore, from the many band structure features, pinpointing those responsible for the anomalous responses is highly desirable.

\begin{figure}
    \centering
    \includegraphics[width=\linewidth]{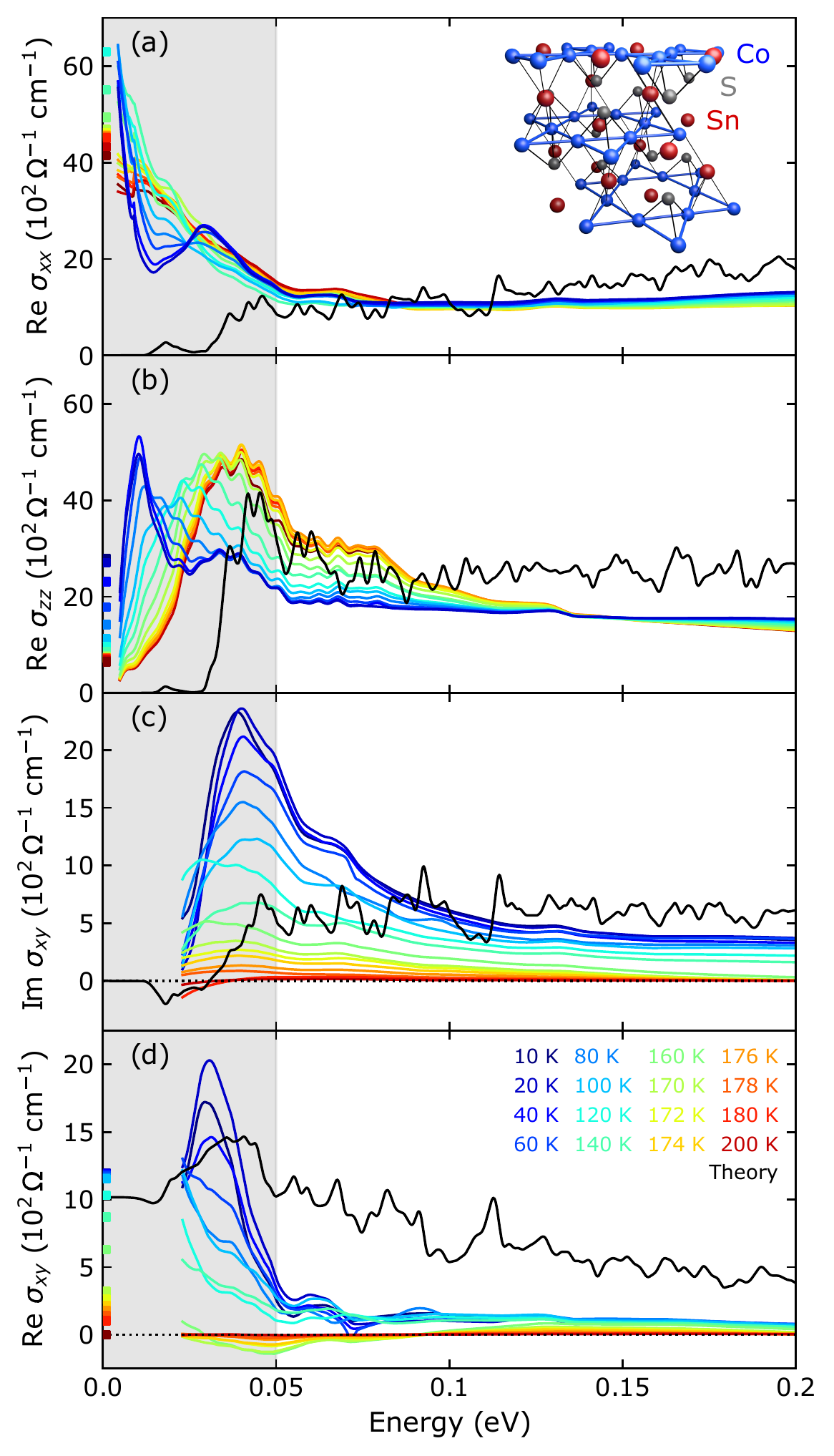}
    \caption{Comparison of the experimental conductivity spectra measured between 10 and 200\,K (colored lines) and the theoretical DFT spectra (black lines) calculated as described in the text. a), b), c) and d) respectively show the real parts of the diagonal, $\Re\,\sigma_{xx}$ \& $\Re\,\sigma_{zz}$, as well as, the imaginary and real part of the off-diagonal conductivity spectra, $\Im\,\sigma_{xy}$ and $\Re\,\sigma_{xy}$. For comparison, the static conductivity values are plotted as colored squares at zero energy.}
    \label{fig:exp_theory}
\end{figure}

This is an especially important question in the prototypical magnetic Weyl semimetal Co$_3$Sn$_2$S$_2$ with individual Weyl points remaining degenerate from SOC gapped nodal loops. Its crystal structure (space group $R\bar{3}m$) consists of an $ABC$-type stack of Co$_3$Sn kagome layers (see inset in Fig.~\ref{fig:exp_theory}(a)), and belongs to the family of shandites \cite{Zabel1979}. Below $T_c=177\,$K, a ferromagnetic order develops with the moments aligned towards the $c$ axis at low temperatures. Just below the transition, an anomalous magnetic phase with non-collinear order was proposed \cite{Kassem2017, Guguchia2020}, but more recent experiments suggest that the domain configuration changes instead \cite{Lee2022, Soh2022}. Due to its kagome structure, Co$_3$Sn$_2$S$_2$ possesses nontrivial electronic topology. In addition to flat bands \cite{Yin2019,Xu2020}, non-relativistic density functional theory (DFT) calculations propose nodal loops on high-symmetry planes of the BZ, which are gapped upon including SOC, each leaving behind a pair of Weyl nodes \cite{Wang2018,Minami2020}. Angle-resolved photoemission spectroscopy (ARPES) studies confirmed the existence of Fermi arcs in this system \cite{Liu2019} and chiral edge modes were found in scanning tunneling microscopy (STM) \cite{Howard2021}. The Berry curvature accumulated by the anticrossing line is claimed to be the source of large anomalous Hall and Nernst effects in this material \cite{Wang2018,Liu2018,Minami2020}. The magnitude of the former reaches as high as $1200\,\Omega^{-1}$cm$^{-1}$ and is therefore comparable to the AHE in the related compound Fe$_3$Sn$_2$ \cite{Wang2016,Ye2018}, in which, as demonstrated recently, only a fraction of the intrinsic AHE can be attributed to twisted nodal lines \cite{Schilberth2022}. In addition, although the DFT band structure of Co$_3$Sn$_2$S$_2$ is relatively simple close to the Fermi level, it is to date unclear which band structure feature dominates the AHE: the gapped nodal loop or the Weyl points.

Here, we address this fundamental question and determine the full optical conductivity tensor of Co$_3$Sn$_2$S$_2$ by polarized infrared reflectivity and magneto-optical Kerr effect (MOKE) spectroscopy (For simplicity and readability, we omit the explicit frequency dependence for the rest of the paper):
\begin{equation}
		\hat{\sigma}(\omega)=\left(\begin{matrix}
		\sigma_{xx}(\omega) & \sigma_{xy}(\omega) & 0\\
		-\sigma_{xy}(\omega) & \sigma_{xx}(\omega) &0 \\
        0 & 0 & \sigma_{zz}(\omega)
		\end{matrix}\right).
\end{equation}
Our results indicate that linear dichroism, namely the ratio of the conductivity in the kagome plane, $\sigma_{xx}$ and out-of-plane, $\sigma_{zz}$, is a sensitive probe of topological features of layered materials. We find a resonant enhancement of the linear dichroism due to transitions along the gapped nodal line. We reveal a giant magneto-optical optical activity in the same energy range by extending MOKE spectroscopy down to $\hbar\omega=25$\,meV. Specifically, we observe 1) a resonance peak at 40\,meV in the Hall conductivity spectrum, $\sigma_{xy}$, which has not been detected before, and 2) directly capture the fingerprints of the nodal loop in the optical conductivity without any extrapolation of $\sigma_{xy}$. Complemented by \textit{ab-initio} calculations, we analyse the momentum space distribution of the Hall spectral weight, which allows us to disentangle the contributions of the gapped nodal line and Weyl points.

Polarized reflectivity spectra were measured on polished $ab$ and $ac$ surfaces of single crystals with a lateral size of $\sim$5\,mm and $\sim$3\,mm, respectively over the range of 0.01 -- 3\,eV. The diagonal optical conductivity spectra were obtained by Kramers-Kronig-transformation of the reflectivity. The magneto-optical Kerr rotation, $\theta$ and ellipticity, $\eta$ were measured at near normal incidence on the same $ab$-cut crystal in $\pm$0.3\,T and were antisymmetrized with respect to the field. The Hall conductivity spectra were calculated using the complex Kerr rotation (0.025 -- 3\,eV) according to
\begin{equation}
    \theta +i\eta=-\frac{\sigma_{xy}}{\sigma_{xx}\sqrt{1+i\frac{1}{\varepsilon_0\omega}\sigma_{xx}}},
    \label{eq:Kerr_effect}
\end{equation}
where $\omega$ is the angular frequency of the photon and $\epsilon_0$ is the vacuum permittivity \footnote{The experimental and theoretical details are presented in the Supplemental Material.}.


The reflectivity and MOKE spectra measured between 10-200\,K are included in Fig.~\ref{fig:measurables}. The out-of-plane reflectivity  remarkably differs from the in-plane spectrum, the latter being in agreement with earlier reports \cite{Xu2020,Yang2020}. In the overlapping energy range, the MOKE spectra agree with those published in Ref.~\onlinecite{Okamura2020}. Importantly, our Kerr rotation and ellipticity obey the Kramers-Kronig relation and fulfil the magneto-optical sum rule, requiring that both parameters approach 0 for $\omega\rightarrow0$.

We show the low-energy spectrum for each independent component of the optical conductivity tensor in Fig.~\ref{fig:exp_theory} (for a broad energy range see Fig.~\ref{fig:cond_broad}). The corresponding static conductivity values are shown for comparison and agree well with the respective spectra at the low-energy cutoff. In panel (a), the real part of $\sigma_{xx}$ exhibits a Drude-like increase towards zero energy responsible for the static conductivity. At 30\,meV, a peak is forming below 100\,K, separated well from the free carrier response. For even higher energies, we find a small temperature dependent hump around 0.25\,eV and a step edge around 0.6\,eV before the conductivity becomes flat without significant temperature dependence. These features agree with earlier reports \cite{Yang2020,Xu2020,Okamura2020}. 

The out-of-plane conductivity spectrum, $\sigma_{zz}$ in panel (b), strongly deviates from $\sigma_{xx}$. Most strikingly, no sign of a Drude peak is observed down to our low-enegy cutoff, and the dc conductivity is also much lower for this direction, indicated by the colored points at zero energy. Therefore, we suspect that within the kagome planes, the strong orbital overlap between Co-sites can produce a coherent conduction, manifested in the Drude term. In contrast, the transport is likely due to incoherent hopping perpendicular to the planes \cite{Kezsmarki2006}. At 200\,K, we find a peak at 40\,meV, which shifts to smaller energies upon lowering the temperature until it eventually splits in two below 100\,K. At higher energies, we find a minimum at 0.4\,eV, which becomes sharper at low temperatures, and a step edge at 0.6\,eV. For even higher energies, $\sigma_{zz}$ is featureless though slightly increasing without distinct temperature dependence, similar to $\sigma_{xx}$. 

In Fig.~\ref{fig:exp_theory}(c) \& (d), the Hall conductivity spectrum $\sigma_{xy}$ shows a strong resonance at 40\,meV, in coincidence with the in- and out-of-plane diagonal components and their ratio. We emphasize that this far-infrared range has not been covered so far, while the higher energy part of the spectra agree very well with Ref.~\onlinecite{Okamura2020}. Both its imaginary and real parts exhibit a large enhancement towards low temperatures, where the peak in the real part overshoots the dc-AHE below 60\,K with a magnitude as high as 2000\,$\Omega^{-1}$cm$^{-1}$ at 10\,K. The good agreement between the low-energy tail of the real part of $\sigma_{xy}$ and the dc-AHE together with the formerly published featureless THz data \cite{Okamura2020}, suggests that there are no further excitations in the narrow uncovered energy interval. Furthermore, as the scattering rate obtained from the Drude peak for $\sigma_{xx}$ is below the cutoff for $\sigma_{xy}$, we conclude that the giant anomalous Hall conductivity of Co$_3$Sn$_2$S$_2$ has dominantly intrinsic origin and it is generated by the interband resonance observed here for the first time.

In order to reveal the microscopic origin of the observed spectral features, we performed \textit{ab--initio} calculations providing all symmetry allowed elements of the conductivity tensor \footnote{The obtained band structure along general high symmetry points is shown in Fig.\,\ref{fig:high_sym_bands}(a).}. The theoretical spectra are coplotted with the experiment in Figs.\,\ref{fig:exp_theory} and \ref{fig:cond_broad}. Disregarding the intraband contribution which is not included in the theory, the spectra qualitatively reproduce all conductivity components on a large energy scale, although the spectral features appear slightly shifted to higher energies compared to experiment, which may indicate correlation effects \cite{Xu2020}.

\begin{figure}
    \centering
    \includegraphics[width=\linewidth]{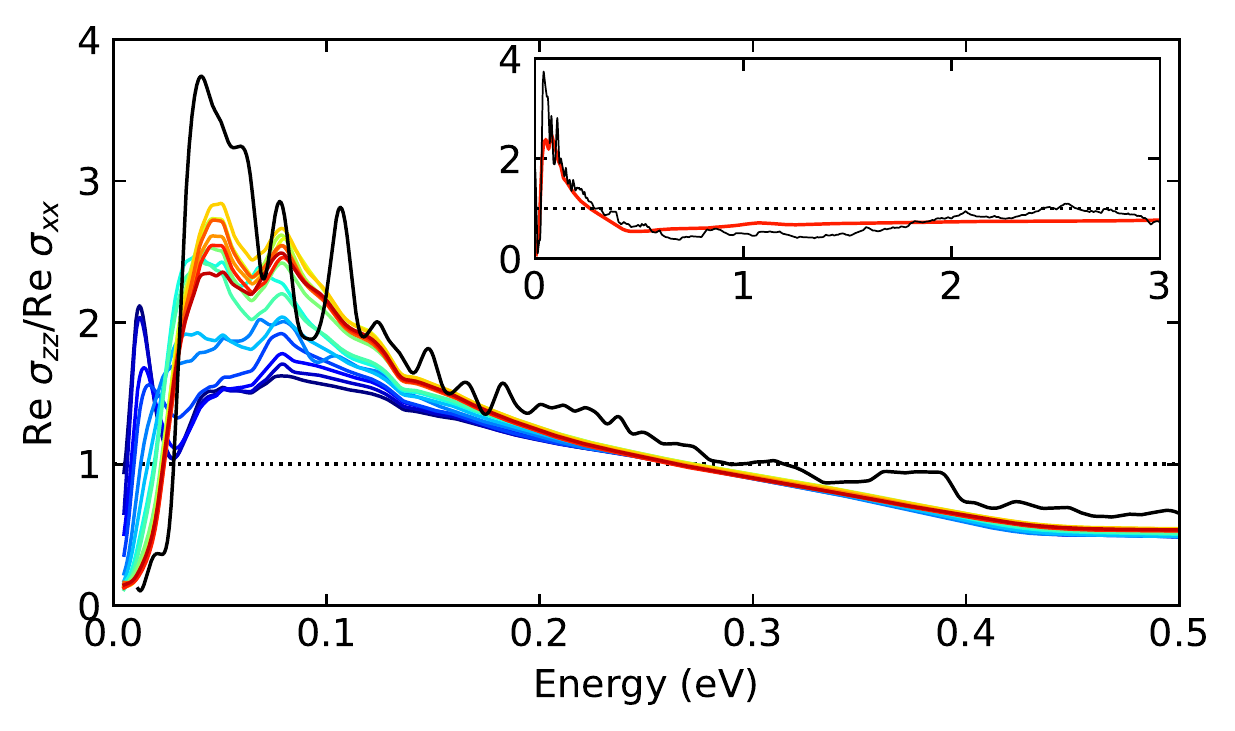}
    \caption{Optical anisotropy spectra $\Re\,\sigma_{zz}/\Re\,\sigma_{xx}$ with the same colorcode as in Fig.\,\ref{fig:exp_theory}. The inset shows the spectra over a broader energy range.}
    \label{fig:opt_ani}
\end{figure}

We quantified the linear dichroism for in- and out-of-plane polarizations by calculating $\Re$\,$\sigma_{zz}/\Re$\,$\sigma_{xx}$ from the spectra shown in Fig.~\ref{fig:exp_theory}. As demonstrated by Fig.\,\ref{fig:opt_ani}, the theory agrees with the experiment remarkably well, especially at higher temperatures. We find that around the resonance at 40\,meV, $\sigma_{zz}$ becomes three times larger than $\sigma_{xx}$. As the temperature is decreased in the experiment, some of the spectral weight splits and moves to lower energies. Although this is not properly captured by the DFT calculations the overall tendency, that $\sigma_{zz}$ is stronger at the resonances, remains valid.

\begin{figure*}
    \centering
    \includegraphics[width=\linewidth]{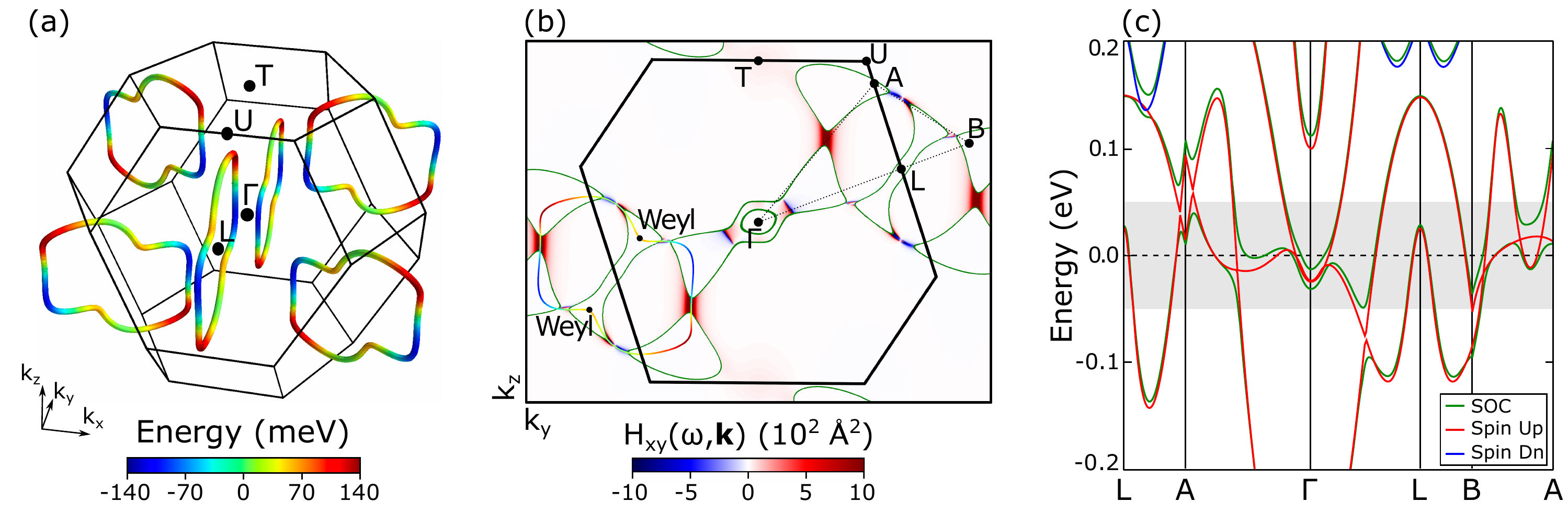}
    \caption{(a) Location of the nodal lines in the Brillouin zone. The colorscale encodes the position relative to the Fermi energy. (b) The Hall spectral weight of the calculated 35.9\,meV peak on the mirror plane containing the nodal line and Weyl points. (c) Bandstructure along the main contributing areas of the real part of the off-diagonal optical conductivity. The grey shading highlights the energy range below 50\,meV where we expect contributions to the peak in $\Re\,\sigma_{xy}$. The non-high-symmetry points $A$ and $B$ are (0.0, 0.4002, 0.301) and (0.0, 0.7373, 0.0) in reciprocal lattice units.}
    \label{fig:kslice}
\end{figure*}

The Hall conductivity is also well reproduced by the theory as shown in Figs.\,\ref{fig:exp_theory} and \ref{fig:cond_broad}(c) \& (d). The imaginary part shows a sudden increase at 40\,meV, where the resonance is observed in the experiment, though it is not that pronounced. Importantly, the theory properly captures the 35.9\,meV peak in $\Re \sigma_{xy}$. The experimental feature is somewhat sharper like in the case of the imaginary part, which may again be related to electronic correlations \cite{Xu2020}. The dc--extrapolation of the theory yields the same AHE as the magnetotransport measurements. Backed with this remarkable agreement, we now analyse the band structure origin of this giant optical Hall effect signal.


We deduced the momentum decomposition of the off-diagonal conductivity by introducing the Hall spectral weight $H_{xy}(\omega,\mathbf{k})$ as $\sigma_{xy}(\omega)=ie^2/\hbar V\int_{BZ}H_{xy}(\omega,\mathbf{k})$. (For details see Eq.\,\ref{eq:Hall-weight} and text of SM). Fig.\,\ref{fig:kslice}(b) shows the Hall spectral weight of the 35.9\,meV low-energy peak in $\Re\,\sigma_{xy}$ on one of the mirror planes containing the nodal line and the Weyl points as indicated by the rainbow line and black labels, respectively \cite{Minami2020,Kanagaraj2022}. The black line is the border of the BZ, the green lines plot the Fermi surface, and the colormap presents the spectral weight. For better orientation, panel (a) plots the nodal lines where the coloring encodes the position of the band crossing relative to the Fermi energy.

We find several hot spots of Hall conductivity on the slice. Interestingly, there is no significant contribution from the Weyl points as they are located 60\,meV above the Fermi energy and cannot contribute to the response in the range below \cite{Minami2020,Kanagaraj2022}. Instead, we find regions close to $\Gamma$ and in the vicinity of $A$ with positive and negative weights right next to each other. Since the conductivity is determined by integration over the BZ, the opposite weight from these points cancel to a large extent. This leaves the large positive patch on the $\Gamma-A$ line giving the dominant contribution. Aside from the regions around $\Gamma$, we always observe a hot spot when the nodal line comes close to or crosses the Fermi energy (light green segments of the loop), hence the hotspots always connect two close regions of the Fermi surface (dark green lines).

Although the hot spots seem to have a similar origin, they behave differently in producing optical weight. In order to see the underlying shape of the bands, we introduce two non-high-symmetry points $A$ and $B$, to plot the band structure along cuts through the hot spots, shown in Fig.\,\ref{fig:kslice}(c) with and without SOC. The grey shading indicates the relevant low-energy range and is shaded in the same fashion in Fig.\,\ref{fig:exp_theory}. Along the $B-A$ line, we cut through a hot spot with subsequent positive and negative weights. Here, the two red spin up bands forming the nodal line are close to the Fermi energy, so with SOC, one of the bands is filled and the other is empty in between the crossings, allowing the optical transitions. Importantly, the tilt of the crossing points is opposite, which is the reason for the different signs of their contribution to $\sigma_{xy}$ \cite{Steiner2017,Sonowal2019}. Around $\Gamma$, we have a similar situation where two red spin up band inversions appear with opposite tilt. The SOC opens a gap, which allows transitions once the upper band is above the Fermi energy, also producing an optical Hall response.

The situation must be different along the $A-\Gamma$ line, where the large positive hot spot does not have a negative partner nearby. Here, we observe only one strongly tilted crossing. Again, with SOC the upper band is pushed above the Fermi level enabling the transition. Due to the strong tilt, the two bands stay nearly parallel over a relatively broad $k$-interval, so the Berry curvature of the gapped nodal line is summed up over a narrow energy range which produces the large patch of optical weight. By inversion and 3--fold rotation symmetry, we expect a total of 6 such spots in the BZ dominating the anomalous Hall conductivity.

Since the interband transition from this feature also gives rise to diagonal conductivity, we directly compare $\sigma_{xx}$ and $\sigma_{xy}$ by calculating the Hall angle, $\Theta_\text{H}=\text{arctan}\Re(\sigma_{xy}/\sigma_{xx})$, which is shown in Fig.\,\ref{fig:ani_Hall}(b). At 40\,meV, the giant off--diagonal conductivity has almost the same magnitude as the diagonal conductivity producing a very large Hall angle of 42.7$^\circ$. When the Kubo formula is written using the circular momentum operators

\begin{multline}
\label{eq:Im_sxy}
\text{Im~}\sigma_{xy}=\frac{e^2\pi}{4m^2V\hbar}\sum_{\mathbf{k},n,n^\prime}\frac{\left|f(\varepsilon_n(\mathbf{k}))-f(\varepsilon_n^\prime(\mathbf{k}))\right|}{\omega_{nn^\prime}}\delta(\omega-\omega_{nn^\prime})\\
\times \left[\left|\bra{n,\mathbf{k}}p_+\ket{n^\prime,\mathbf{k}}\right|^2-\left|\bra{n,\mathbf{k}}p_-\ket{n^\prime,\mathbf{k}}\right|^2\right],
\end{multline}
$\sigma_{xy}$ depends on the difference of the two circular components, whereas $\sigma_{xx}$ is given by their sum \cite{Antonov2004}. As a consequence, the largest possible Hall angle for an interband transition is 45$^\circ$. This is realized when one matrix element is zero, which indicates an almost fully polarised transition in the present case, yielding a nodal line resonance.

In summary, we provide a showcase for disentangling the contributions of various topological features to the AHE, and identify the gapped nodal line as the main source of the giant AHE in Co$_3$Sn$_2$S$_2$. Facilitated by far infrared MOKE spectroscopy, we observe a low-energy magneto-optical resonance for the first time. As the zero energy extrapolation of this interband transition explains the static Hall conductivity, we confirmed that the AHE is dominantly intrinsic. Our \textit{ab-initio} calculations are able to reproduce the experimental spectra with remarkable accuracy, allowing a momentum and band decomposition of the optical Hall conductivity. We find that the Weyl points located 60\,meV above the Fermi energy only yield singular contributions in a small $k$-volume. By contrast, the nodal line segments approaching or crossing the Fermi energy produce large AHE hotspots after being gapped by SOC. In addition, we verify that the tilt of the nodal line is a crucial factor, which can lead to pairwise cancellation or in contrast, produce a nodal line resonance. Remarkably, the linear dichroism is significantly enhanced by the nodal line resonance, leading to a potentially new signature of topological states. Finally, we note that this magneto-optical analysis is applicable for any material where large AHE is suspected from topological bands. Since in magnetic materials, the electronic topology may couple to the magnetic order, it is also a suitable tool to monitor the effects of e.g.\,external fields for manipulating topological properties.

\begin{acknowledgements}
The authors are grateful to Christine Kuntscher, Liviu Chioncel, and Artem Pronin for fruitful discussions. This work was supported by the Hungarian National Research, Development and Innovation Office – NKFIH grants FK 135003 and Bolyai 00318/20/11 and by the Ministry of Innovation and Technology and the National Research, Development and Innovation Office within the Quantum Information National Laboratory of Hungary. Sándor Bordács is supported by the ÚNKP-22-5-BME-280 new national excellence program of the ministry for innovation and technology from the source of the national research, development and innovation fund.
\end{acknowledgements}

\bibliography{References}

\clearpage
\section*{Supplemental Material}
\beginsupplement
\subsection*{Crystal growth for the magneto-optical measurements}
A large single crystal of Co$_3$Sn$_2$S$_2$ was grown in a cylindro-conical shape (1\,cm in diameter and 5\,cm in length) from its melt by a modified Bridgman method \cite{Holder2009,Kassem2015}. About 10\,g of polycrystalline Co$_3$Sn$_2$S$_2$ synthesized by a solid state reaction was charged in a tipped glassy carbon crucible which was sealed under vacuum in a quartz tube. The sealed ampule was suspended by a Kanthal thread from the top to the hot zone of a vertical tube furnace and heated over 30 hours up to $1000^\circ$C, kept for 6\,h, and then slowly cooled over 72 h to $800^\circ$C. After furnace cooling, a crystal has been removed from the crucible and could be easily cleaved in the (001) plane. Crushed parts were investigated by powder x-ray diffraction (XRD), Laue x-ray spectroscopy and wave-length dispersive x-ray spectroscopy (WDX), those indicated a single-phase and high-quality grown crystal with stoichiometric chemical composition of Co$_3$Sn$_2$S$_2$.

\subsection*{Crystal growth for magnetotransport}
Single crystals of Co$_3$Sn$_2$S$_2$ were grown, in a flat hexagonal shape, out of Sn flux as published elsewhere \cite{Kassem2020}. Lumps of Co (99.9\,\% high purity chemicals) and grains of Sn (99.999\,\% high purity chemicals) were mixed with S powder (99.999\,\% nacalai tesque) in a molar ratio of  Co : S: Sn = 8: 6: 86. A mixture with a total mass of $\approx15\,$g was charged in Al$_2$O$_3$ growth crucible.  The growth crucible was covered by another inverted Al$_2$O$_3$ crucible filled with quartz wool and both were sealed under vacuum in a quartz tube. The ampule was fired in a muffle furnace at $1050^\circ$C for 6 hours and then slowly cooled to $700^\circ$C over 72 hours at which the ampule was removed from the furnace. The flux was removed from the crystals via a rapid decanting of the ampule followed by a subsequent spinning in a centrifuge. The grown crystals were characterized by powder XRD and WDX and the crystal orientations were determined by Laue x-ray spectroscopy.

\subsection*{Reflectivity measurements}
The reflectivity spectra of Co$_3$Sn$_2$S$_2$ were obtained in a Bruker IFS/66 FTIR-spectrometer for the MIR-VIS range and a Bruker Vertex 80v for the FIR. The spectra were measured in the frequency range 80-32000\,cm$^{-1}$ (0.01-4\,eV) from room temperature down to 10\,K. As references, a silver and gold mirror where used in the MIR-VIS and FIR experiments, respectively. The optical conductivity $\sigma_1$ was calculated by using Kramers-Kronig analysis on the merged spectra. At this point, the low-energy side of the reflectivity spectrum was extrapolated by using a Hagen-Rubens law and the dc-conductivity, while for the UV the reflectivity spectrum was extrapolated with free electron behaviour setting in at $10^6$\,cm$^{-1}$ and an exponent for the interband regime of 1.5.

\subsection*{MOKE-spectroscopy}
The broadband MOKE spectra were recorded in near-normal incidence and were combined from several measurements in different frequency ranges, employing grating and interferometer based spectrometers as described elsewhere \cite{Sato1981, Demko2012, Bordacs2010, Schilberth2022}. Small permanent magnets provided a field of 0.3\,T at the sample position. Because of the large uniaxial anisotropy in this material, the sample was warmed up over $T_c$ between measurements with reversed field direction to ensure proper antisymmetrisation.

In the overlapping energy range, the MOKE spectra agree with those published in Ref.~\onlinecite{Okamura2020}. Beside the giant Kerr rotation with a peak magnitude of -3.3\,deg at 0.09\,eV, we resolve a peak around 50\,meV in the ellipticity with a magnitude of 2\,deg, which was not detected before.

\subsection*{DFT calculations}
The electronic structure of Co$_3$Sn$_2$S$_2$ is calculated using VASP code \cite{Kresse1996a,Kresse1996b,Kresse1999} based on the density functional theory. The generalized gradient approximation of Perdew-Burke-Ernzerhof was adopted for the exchange-correlation functional \cite{Perdew1996}. The crystal structure for Co$_3$Sn$_2$S$_2$ is of trigonal form with $a$ = 5.379\,\AA~and $\alpha$ = 59.8658$^\circ$, which is the experimentally determined lattice constant \cite{Vaqueiro2009}. In the self-consistent band structure calculations, $\Gamma$-centered $k$ meshes of $24\times 24 \times 24$ were used in the Brillouin zone integration. The optical properties are further evaluated using the Wannier functions, and the Kubo-Greenwood formula \cite{Pizzi2020}. 

Below, we present the formula used for calculating Fig.  1(c) and deducing the Hall spectral weight of a certain transition energy plotted in Fig. 3(b).

\begin{align}
\label{eq:Hall-weight}
\sigma_{\textbf{\textit{k}},\alpha\beta}(\hbar\omega)&=\frac{i e^2}{\hbar V}\sum_{\textbf{\textit{k}},n,n^\prime}(f_{n^\prime,\textbf{\textit{k}}}-f_{n,\textbf{\textit{k}}})Re\left[\frac{\epsilon_{n^\prime,\textbf{\textit{k}}}-\epsilon_{n,\textbf{\textit{k}}}}{\epsilon_{n^\prime,\textbf{\textit{k}}}-\epsilon_{n,\textbf{\textit{k}}}-(\hbar\omega+i\eta)}\right]\nonumber \\
&A_{nn^\prime,\alpha}(\textbf{\textit{k}})A_{nn^\prime,\beta}(\textbf{\textit{k}})\nonumber \\
&=\frac{ie^2}{\hbar V}\sum_{\textbf{\textit{k}}}H_{\alpha\beta}(\omega,\textbf{\textit{k}})
\end{align}
The $\alpha$ and $\beta$ are the indices in Cartesian coordinates, $V$ is the cell volume, and $f_{n,\textbf{\textit{k}}} = f(\epsilon_{n,\textbf{\textit{k}}})$ is the Fermi-Dirac distribution function. $\omega$ is the optical frequency and $\eta>0$ is the smearing parameter. 

Finally, $A_{nn^\prime,\alpha} = \left\langle u_{n,k} \middle | i \nabla_{k_\alpha} \middle |u_{n^\prime,k} \right\rangle$ is the Berry connection and $H_{\alpha\beta}$ is the Hall spectral weight at certain transition energy. Importantly, if we apply $\left\langle v\right\rangle=\left\langle p \right\rangle/m$ and the relation $\left\langle \psi_{n,k}\middle|v \middle | \psi_{n^\prime,k} \right\rangle =-i/\hbar\cdot(\epsilon_{n^\prime,\textbf{\textit{k}}}-\epsilon_{n,\textbf{\textit{k}}})A_{nn^\prime}(k)$, we would arrive at Eq.\,\ref{eq:Im_sxy} of the main text. If we take $\omega$ and $\eta$ as zero, we obtain the ordinary Berry curvature for the anomalous Hall effect.

A fine mesh of 100 $\times$ 100 $\times$ 100 $k$ points are applied during the integration with good convergence. The Wannier functions were constructed using Co $d$, Sn $s$, $p$, and S $s$, $p$ orbitals with a resulting tight-binding model well describing the DFT band structures. Importantly, from the main text we notice that the experimental low-frequency peak of the real part off-diagonal optical conductivity $\Re \sigma_{xy}$ is nicely captured by the theoretical calculations (Fig. \ref{fig:exp_theory}(d)).

\begin{figure}
    \centering
    \includegraphics[width=\linewidth]{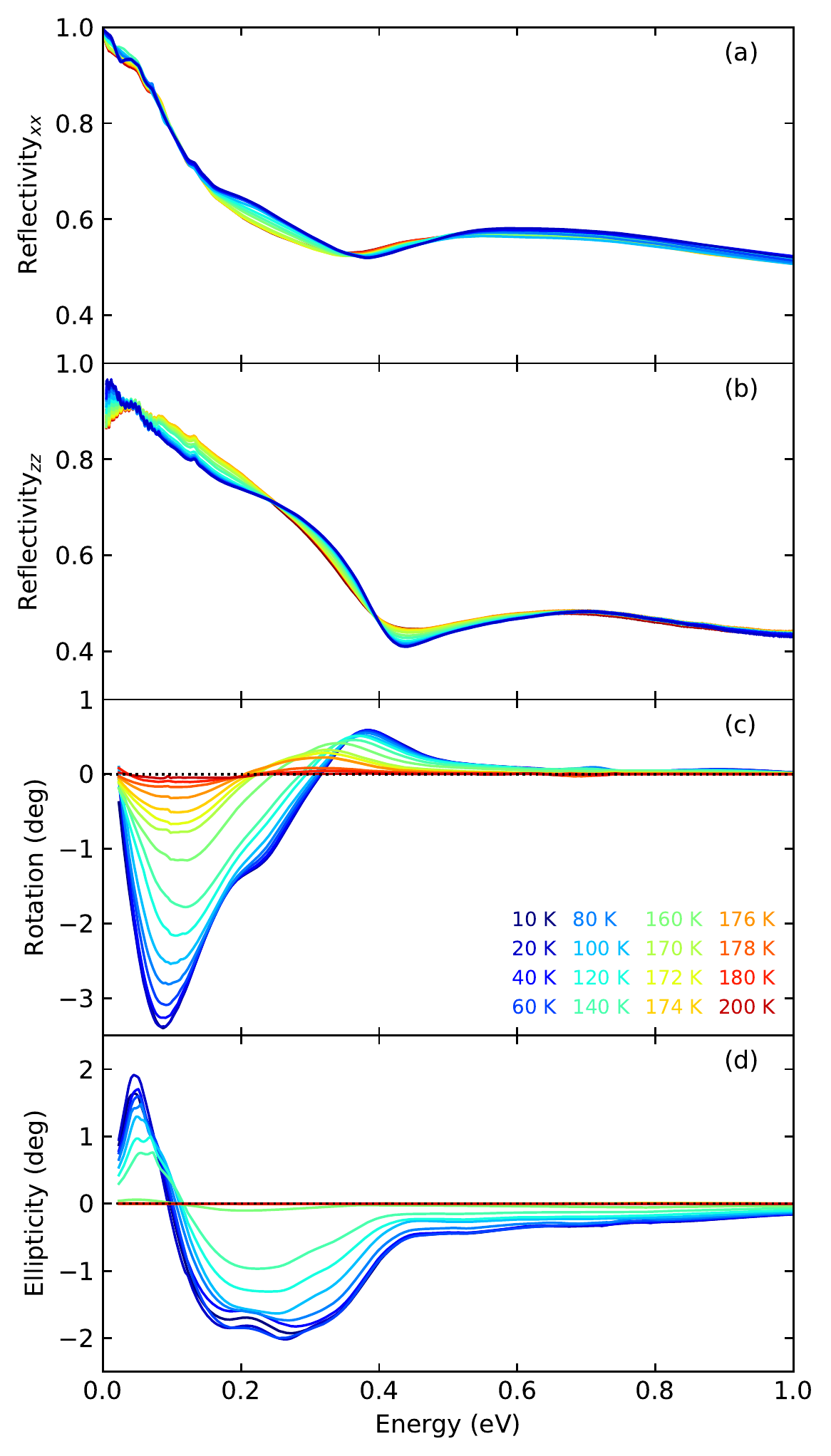}
    \caption{Spectra measured for several temperatures between 10 and 200\,K of a) the reflectivity on the kagome plane and b) along the stacking direction c) Kerr-rotation and d) ellipticity on the $ab$-plane in the energy range up to 1\,eV.}
    \label{fig:measurables}
\end{figure}

\begin{figure}
    \centering
    \includegraphics[width=\linewidth]{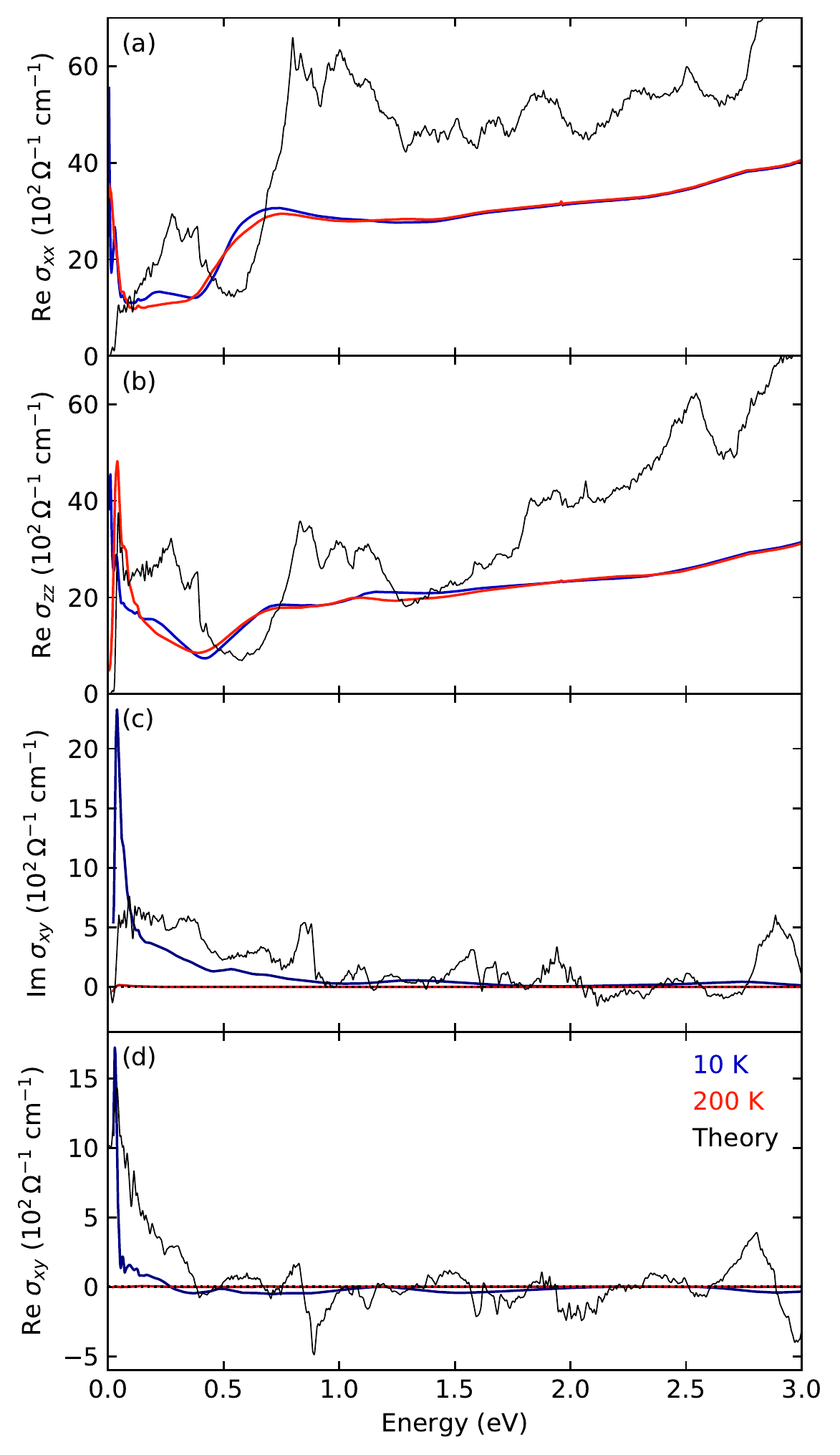}
    \caption{Conductivity spectra over a broad energy range.}
    \label{fig:cond_broad}
\end{figure}

\begin{figure*}
    \centering
    \includegraphics[width=\linewidth]{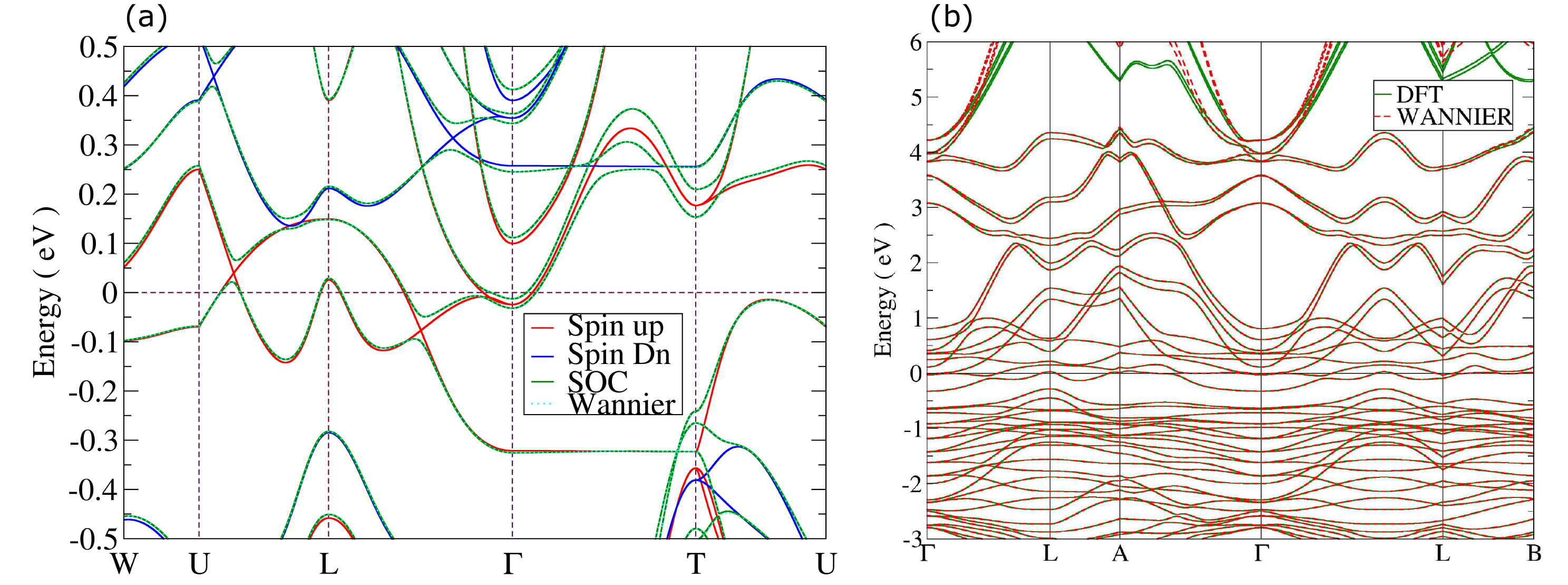}
    \caption{a) Calculated band structures along the high symmetry line with and without the spin-orbit coupling. The eigenvalues of the Wannier tight-binding model are also plotted. b) Comparison of band structures over a large energy range calculated by density functional theory and Wannier tight-binding model.}
    \label{fig:high_sym_bands}
\end{figure*}

\begin{figure*}
    \centering
    \includegraphics[width=0.5\linewidth]{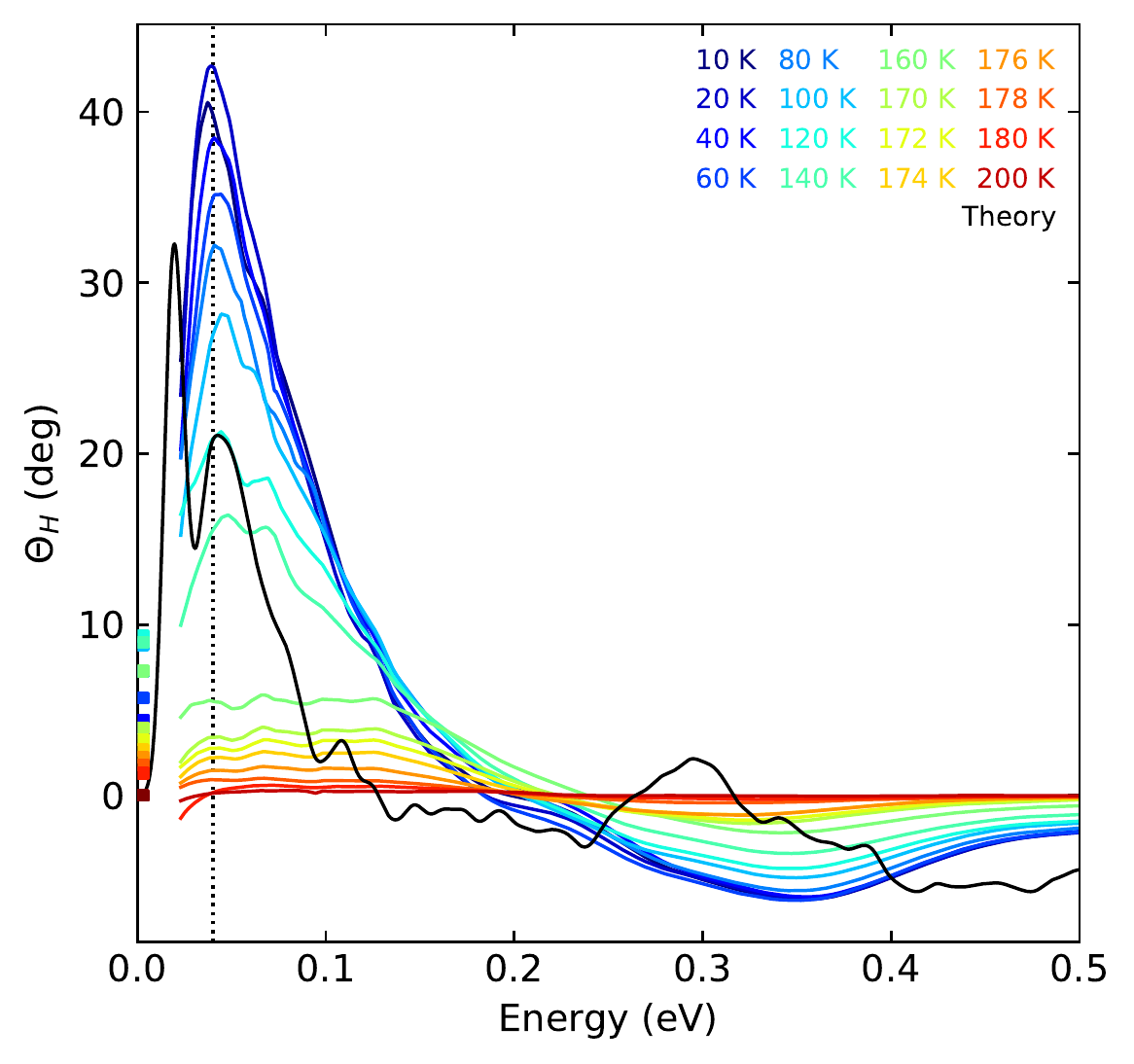}
    \caption{Hall angle spectra $\text{arctan}\Re(\sigma_{xy}/\sigma_{xx})$ with a maximum of 42.7$^\circ$.}
    \label{fig:ani_Hall}
\end{figure*}

\end{document}